# 25 Tweets to Know You: A New Model to Predict Personality with Social Media

**Pierre-Hadrien Arnoux, Anbang Xu, Neil Boyette, Jalal Mahmud, Rama Akkiraju, Vibha Sinha**

IBM Research - Almaden, San Jose, CA, USA

{parnoux, anbangxu, neil.boyette, jumahmud, akkiraju, vibha.sinha}@us.ibm.com

**Abstract**

Predicting personality is essential for social applications supporting human-centered activities, yet prior modeling methods with users' written text require too much input data to be realistically used in the context of social media. In this work, we aim to drastically reduce the data requirement for personality modeling and develop a model that is applicable to most users on Twitter. Our model integrates Word Embedding features with Gaussian Processes regression. Based on the evaluation of over 1.3K users on Twitter, we find that our model achieves comparable or better accuracy than state-of-the-art techniques with 8 times fewer data.

## Introduction

There is a growing trend in social applications to consider users' personality to provide more adaptive and personalized user experience (Hu, et al. 2016; Liu, et al. 2016). Users' self-authored text is often used to predict personality measured by the Big-5 personality dimensions, Openness, Conscientiousness, Extraversion, Agreeableness, Neuroticism (OCEAN) (McCrae and John 1992; Schwartz, et al. 2013; Yarkoni 2010). With hundreds of millions of users participating on social media and sharing self-authored content, social media provides a tremendous opportunity for personality modeling.

However, prior modeling methods require too much input data to be realistically used in the context of social media. Indeed, prior works report modeling accuracy using on average 200 Facebook posts (Schwartz, et al. 2013) or even 100 000 words (Yarkoni 2010). In contrast, users have on average 22 posts on Twitter (Burger, et al. 2011). Hence, it is unclear how well these models would work with real life scenarios, where a majority of users have small amounts of text for analysis.

This paper studies the accuracy of prior works on Big-5 personality inference as a function of the size of the input text. Moreover, we propose a new method for personality inference, which significantly reduces the text size requirement. Figure 1. shows that the proposed method outperforms the state-of-the-art methods overall and requires 8 times fewer data to perform at the same level, making it applicable to a wider population who uses social media platforms like Twitter.

The main contributions of this paper are as follows:

- We present the first study of personality prediction on a small amount of text to see how our method and previous ones perform in a real life social media context.
- We introduce the use of Word Embedding as features for personality modeling and Gaussian Processes as the learning algorithm. This method outperforms previous works in the field.

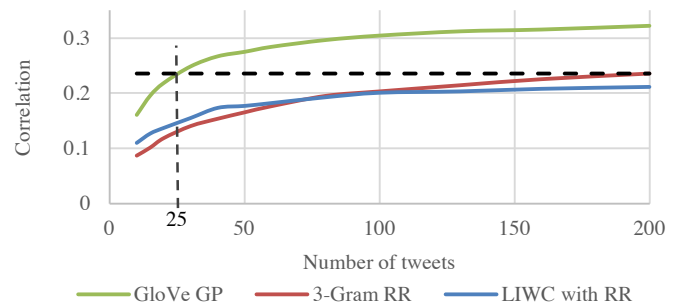

*Figure 1. Prediction accuracy of the Big-5 traits according to the number of tweets. Reported correlations are significant $p < 0.01$.*

## Word Embedding with Gaussian Processes

Our method combines Word Embedding with Gaussian Processes. We extract the words from the users' tweets and average their Word Embedding representation into a single vector. The Gaussian Processes model then takes these vectors as an input for training and testing.

**The Word Embedding features**

In this study, we introduce Word Embedding features to the field of personality modeling. Word Embedding is a technique that represents words as a dense, low-dimensional and real-valued vector. It relies on syntactic and semantic relationships between words. Usually learned from large amounts of unstructured text data, this representation helps learning algorithms achieve better results on natural language processing tasks by bringing similar words closer together (Mikolov and Dean 2013). It was also shown to improve learning methods dealing with short text (Kenter and de Rijke 2015).

While there are many available Word Embedding models, we chose to use the Twitter 200 dimensional GloVe model (Pennington, et al. 2014) as it was trained on 2B tweets.

**The Gaussian Processes model**

This work also introduces a new non-linear model: Gaussian Processes (GP) (Rasmussen 2006). Used in natural language processing and data mining, GP is very well suited for regression as it allows an explicit quantification of noise and a modulation of features usefulness via fitting a kernel function to the data. Recent works have shown that GP combined with Word Embedding can be very efficient in short text classification (Ma, et al. 2015) or in nonlinear modeling from text features (Yoshikawa, et al. 2015).

For our method, we use the 200 dimensional vector from our Word Embedding feature generation step as an input and train a GP model for each of the Big-5 dimensions.

## Experiment design

We obtain the groundtruth data by surveying over 1.3K participants, and we compare the performance of our method with previous methods under various settings.

**Ground Truth Collection**

Following previous work (Schwartz, et al. 2013), a survey is conducted to collect participants' self-reported personality ratings as well as their tweets. To reach a participant, we create a web application and we advertised it via Twitter adds. Through this app, participants voluntarily agree to share their tweets and answer a personality survey. The survey adopts the 50-item International Personality Item Pool form (Goldberg, et al. 2006) to assess participants' Big-5 personality traits and takes about 15 minutes to complete. Once the survey is completed, participants are presented with their personality scores. In this setting, participants are motivated to complete the survey in order to view their personality assessments.

We recruit 1323 participants with at least 200 non re-tweet tweets ($\mu$ = 1020, $\delta$ = 541). This enables us to design a sampling experiment with numbers of tweets from 10 to 200 and have the same user set for all experiments.

The age distribution of the participants is: under 18 (23%), 18 to 24 (47%), 25 to 34 (14%), 35 to 54 (12%), above 54 (3%). 52% of the participants are female.

We normalize the survey score to have them between 0-1. The distribution of score is the following: O($\mu$ = 0.76, $\delta$ = 0.12), C($\mu$ = 0.59, $\delta$ = 0.15), E($\mu$ = 0.54, $\delta$ = 0.18), A($\mu$ = 0.72, $\delta$ = 0.13), N($\mu$ = 0.44, $\delta$ = 0.19). This is consistent with previous work (Golbeck, et al. 2011).

To pre-process the tweets, we use the followings steps: we remove the URLs and the hashtags; we set the text to lowercase, and remove the numbers and the punctuations.

**Methods for Comparison**

We compare our method with two state-of-the-art methods:

*Linguistic Inquiry and Word Count (LIWC) with Ridge Regression (RR)*. The original method proposed by (Yarkoni 2010) uses LIWC (Pennebaker, et al. 2015) to extract features and RR as a learning algorithm.

*3-Gram with Ridge Regression*. We implement the previous state-of-the-art method (Schwartz, et al. 2013) by using 3-Gram and RR.

*Word Embedding with Gaussian Process (GP)*. Our method integrates GloVe features with Gaussian Process regression as the learning algorithm.

To tune our parameters and evaluate our performances, we use a 3 sets split: Training, Validation, and Testing. The data is split between Testing and Training using a 10 Fold Cross-Validation and Training is sub-split in Training and Validation using a 75%-25% rule. The performance is evaluated by a Pearson Correlation analysis between the predicted and actual personality scores on the Testing data.

Here we emphasize that both RR and GP are using regularization techniques to reduce overfitting and that all methods are trained and tested on our Twitter dataset.

**Comparison Settings**

The performance of those three methods are compared in three different settings:

*Full Setting.* In this setting, the methods are trained and tested over the entire corpus of texts.

*Sampling Setting.* To simulate users with various numbers of tweets, we perform a downsampling on the tweets of the testing users and vary the number of tweets used.

*Real-life Setting.* The last setting aims at investigating further the performance of the methods in real life applications by training the models on a large population of users with a large number of tweets and testing them on a small set of real life users with small numbers of tweets. In order to perform this analysis, we collect an addition small set of 55 users with no restriction on the number of tweets following the previous procedure. The users had on average 28 non re-tweet tweets (μ = 28, δ = 11), which is in line with the previous number of 22 publically available tweets.

## Results

**Full Setting**

Following prior work, we train and test the models on all the available tweets. In addition, we test 3 other combinations of features and models: GloVe with RR, LIWC with GP and 3-Gram with GP.

Table 1. shows that our method establishes a new state-of-the-art performance with an average correlation of 0.33 over the Big-5 traits, 33% better than the previous best method. Also, consistent with (Schwartz, et al. 2013), we observe that 3-Gram features with RR achieve better results than LIWC features.

From the additional combination tests, it seems that GloVe feature and GP contributes equally to the performance of the method. Indeed, the combination of LIWC GP performs as well as the combination of GloVe RR at an average correlation of 0.26. We also find that GP does not perform well in combination with Bag-Of-Words like features such as 3-Gram, as suggested by (Yoshikawa, et al. 2015).

In addition to this analysis, we are able to compute the coverage of each feature set over our data. GloVe features cover 92% of the words and LIWC 62%.

Limited to the most frequent 2000 unigrams, bi-grams and trigrams, 3-Gram only covers 50% of the unigrams, 25% of the bigrams and 7% of the trigrams. To reach 7% coverage of the trigrams, we generate 750 features and increasing the coverage by 1% would double this number.

**Sampling Setting**

For this analysis, we train our models with all available tweets. Then, for each user in the testing set, we pick 20 random subsets of tweets and test the model on them. We vary the number of tweets in those subsets and report the correlation of the models averaged over the Big-5 traits as a function of the number of tweets.

Figure 1. shows the prediction accuracy given the number of testing tweets. We first confirm that, for large numbers of tweets the performances of the 3 methods are converging towards the results of the previous correlation analysis. We also see that, while for large numbers of tweets the 3-Gram method outperforms the LIWC method, for less than 75 tweets it is the opposite. This highlights the necessity of using non-sparse features when dealing with small texts.

Most importantly, Figure 1. shows that our method outperforms the other methods for all numbers of tweets. For 200 tweets, our method is 37% better than the next best method and with only 25 tweets it is performing on par with the state-of-the-art with 200 tweets, i.e. 8 times fewer data.

**Real-life Setting**

For this real life application example, we report the results of an ANOVA test over the absolute error averaged over the Big-5 traits to establish the significance of the difference we see across methods.

Figure 2. shows the comparative mean absolute error averaged over the Big-5 traits for the set of 55 users. The average absolute error of our method is 25% smaller than the state-of-the-art and 11% smaller than the original method. These results are significant ($F(2,51) = 6.82$, $p < 0.01$).

|  | *Agree.* | *Consc.* | *Extrav.* | *Neurot.* | *Openn.* |
|---|---|---|---|---|---|
| *LIWC RR* | 0.25 | 0.27 | 0.13 | 0.28 | 0.26 |
| *3-Grm RR* | 0.21 | 0.28 | 0.18 | 0.35 | 0.26 |
| *GloVe RR* | 0.27 | 0.27 | 0.27 | 0.27 | 0.27 |
| *LIWC GP* | 0.26 | 0.28 | 0.20 | 0.33 | 0.26 |
| *3-Grm GP* | 0.11 | 0.12 | 0.15 | 0.15 | 0.09 |
| *GloVe GP* | 0.29 | 0.33 | 0.25 | 0.42 | 0.37 |

*Table 1. Model correlation comparison for the Big-5 traits. The reported correlations are significant p<0.01.*

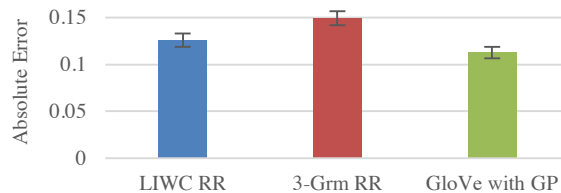

*Figure 2. Mean Absolute Error averaged over the Big-5 traits.*

We also report a pairwise t-test over the same data between our method and the next best method. The two sets are correlated at 0.68 and with a significance of ($p < 0.01$).

These results show that our method outperforms the previous state-of-the-arts in a real life context.

## Discussion

Our first finding is that our proposed method outperforms the state-of-the-art methods for personality prediction in all three experiments. Both Word Embedding features and Gaussian Processes contribute to these performance improvements. As GloVe features are learned over a large corpus of external tweets, its vector representation brings external knowledge to our problem. Also, because of its dense representations of words, our method handles better short texts, as well as unseen data. In addition, the Word Embedding features fit well with GP. Indeed, because of its internal Kernel representation, GP relies on the covariance of the features. If used on features like N-Grams, the Kernel computation fails to capture similarities between documents if similar words don't appear in the same documents. However, when used in combination with GloVe features, and to a lesser extent LIWC features, the Kernel computation is able to rely on embedded co-occurrence of words at a document level.

Our second finding is that LIWC features outperform 3-Gram features on short texts, while 3-Gram features perform better on long texts (75 tweets ≈ 1125 words). This is primarily due to the sparsity of Bag-Of-Words features, and in particular 3-Gram features where one uses groups of at most 3 co-occurring words. Indeed, for our dataset, the most frequent 125 groups of 3 co-occurring words only cover 4% of all groups. Therefore, a short unseen text may not have the same features as the training set.

In this work, we demonstrate that our method is able to predict users' Big-5 personality traits from their social media text in a real-life context. Our method is available as an API (http://bluemix.com) and we believe it can be applied to many social applications to improve user experience.

While this work improves personality modeling, there is a lot of room for improvement in terms of accuracy. Indeed, at its best, our method is only able to achieve an average correlation of 0.33, with the best performance for Neuroticism at 0.42. Future work can further improve this moderate level of correlation. Also, while we investigate the influence of the length of the text used for testing, we only train our model on users with a large number of tweets. Additional studies can be conducted to examine the performance of the method with small numbers of tweets in both training and testing. Finally, our findings are only based on English Twitter data, but we expect this method not to be restricted to one single culture or social media platform. It would be interesting to examine how this prediction model applies to different cultures and platforms.